# Giant bulk spin-orbit torque and efficient electrical switching in single ferrimagnetic FeTb layers with strong perpendicular magnetic anisotropy

Qianbiao Liu[1,2+], Lijun Zhu[1,2+*], Xiyue S. Zhang[2], David A. Muller[2], Daniel C. Ralph[2,3]

1. State Key Laboratory of Superlattices and Microstructures, Institute of Semiconductors, Chinese Academy of Sciences, Beijing 100083, China
2. Cornell University, Ithaca, New York 14850, USA
3. Kavli Institute at Cornell, Ithaca, New York 14850, USA

[+]These authors contributed equally to this work.
*ljzhu@semi.ac.cn

**Abstract:** Efficient manipulation of antiferromagnetically coupled materials that are integration-friendly and have strong perpendicular magnetic anisotropy (PMA) is of great interest for low-power, fast, dense magnetic storage and computing. Here, we report a distinct, giant bulk damping-like spin-orbit torque in strong-PMA ferrimagnetic $Fe_{100-x}Tb_x$ single layers that are integration-friendly (composition-uniform, amorphous, sputter-deposited). For sufficiently-thick layers, this bulk torque is constant in the efficiency per unit layer thickness, $\xi_{DL}^j/t$, with a record-high value of $0.036 \pm 0.008$ nm$^{-1}$, and the dampinglike torque efficiency $\xi_{DL}^j$ achieves very large values for thick layers, up to 300% for 90 nm layers. This giant bulk torque by itself switches tens of nm thick $Fe_{100-x}Tb_x$ layers that have very strong PMA and high coercivity at current densities as low as a few MA/cm$^2$. Surprisingly, for a given layer thickness, $\xi_{DL}^j$ shows strong composition dependence and becomes negative for composition where the total angular momentum is oriented parallel to the magnetization rather than antiparallel. Our findings of giant bulk spin torque efficiency and intriguing torque-compensation correlation will stimulate study of such unique spin-orbit phenomena in a variety of ferrimagnetic hosts. This work paves a promising avenue for developing ultralow-power, fast, dense ferrimagnetic storage and computing devices.

Ferrimagnetic materials can host a variety of exotic properties that are promising for technology, e.g. magnetization/angular momentum compensation, giant perpendicular magnetic anisotropy (PMA)[1-4], ultrafast magnetic domain wall velocities[5-7], reduced sensitivity to stray magnetic fields than ferromagnets, and easier and fast detection than antiferromagnets. Therefore, ferrimagnets (FIMs) are potentially advantageous for dense and fast magnetic recording, memory, and computing applications [8,9]. However, successful integration of FIMs into high-performance functional devices requires efficient manipulation of strong-PMA and integration-friendly FIMs, which has remained a big challenge. So far, electrical switching of FIMs by interfacial spin-orbit torque (SOT)[11] is only reported in samples with poor PMA, small coercivities (e.g. $H_c$ < 0.2 kOe for GdFeCo[3,10,12], $H_c$ <0.25 kOe for CoTb[13,14]), and low SOT efficiencies ($\xi_{DL}^j$ = 0.017 for Pt/CoTb[15]). Moreover, the yet-known bulk SOTs [16-22] in magnetic single layers have low efficiency per unit film thickness [20-27] and/or usually require a single-crystal structure [17] and a composition gradient[16,18,19,22], which limit practical application.

Here, we report a record-strong bulk SOT within single layers of $Fe_{100-x}Tb_x$ alloys (FeTb for short) that have strong PMA, spatially-uniform, amorphous structure, and are grown simply by sputtering onto oxidized silicon substrates, which are highly desirable for large-scale and high-density integration with CMOS circuits. Surprisingly, the efficiency of the bulk torque varies rapidly as a function of stoichiometry and reverses sign in between the compensation points for magnetization and angular momentum. We also show that this bulk SOT can switch tens of nm thick FeTb layers with strong PMA and high $H_c$ at extremely low current densities.

**Sample characterizations**

For this work, we deposit a series of $Fe_{100-x}Tb_x$ single layers with different thicknesses and Tb volume percentages ($x$ = 15-78) by co-sputtering at room temperature. Each sample is capped by a MgO (1.6 nm)/Ta (1.6 nm) bilayer that is fully oxidized upon exposure to the atmosphere [see the electron energy loss spectrum (EELS) results in Ref. 26]. For measurements of SOT and SOT-induced switching, the layers are patterned into Hall bars that are 60 μm long and 5 μm wide. More details about sample preparation and characterization can be found in Supplementary Sec. 1. These FeTb alloys have an amorphous and homogeneous texture and reasonably sharp interfaces, as indicated by cross-sectional scanning transmission electron microscopy (STEM) measurements [see Fig. 1(a) for the example of a $Fe_{57}Tb_{43}$ sample (66 nm)]. Electron energy loss spectroscopy (EELS) measurements demonstrate that there is no obvious variation in the ratio of Fe to Tb concentrations through the thickness of the films (Fig. 1(b)). There is an overall variation in the absolute EELS intensities of Fe and Tb due to variations in thickness along the sample wedge created during the ion-beam thinning process (see Supplemental Sec. 2 and ref. [27]), but this should not be misinterpreted as a composition gradient. There is also no obvious indication of oxidization of the FeTb from the EELS imaging, which is consist with the previous observation[16] that the MgO (1.6 nm)/Ta (1.6 nm) bilayer well protects the magnetic layers from oxidization.

The FeTb samples have strong PMA and square perpendicular magnetization hysteresis loops over a wide composition range, $x$ = 29-61, when the thickness is greater than 7 nm [Fig. 1(c) and Supplementary Fig. S3]. The perpendicular magnetic anisotropy field ($H_k$) is very high and reaches values as large as 100 kOe for 20 nm films near $x$ = 47 [Fig. 1(d)]. In Fig. 1(e)-1(g) we summarize the saturation magnetization ($M_s$), the coercivity ($H_c$), and the anomalous Hall resistance ($R_{AHE}$) for 20 nm $Fe_{100-x}Tb_x$ samples as a function of $x$. As expected, $M_s$ shows a strong, "V-shaped" variation with $x$, which is a characteristic of the competing magnetic moment contributions of the antiparallel Fe and Tb sub-lattices. From the fit of the data to the relation $M_s = (1-0.01x) M_{Fe} – 0.01x\, M_{Tb}$ (see the orange solid line in Fig. 1(e)), we obtain $M_{Fe} = 824 \pm 40$ emu/cm$^3$ ($\approx 1.05\ \mu_B$) and $M_{Tb} = 940 \pm 80$ emu/cm$^3$ ($\approx$ 3.14 $\mu_B$) of our samples. The values of $M_{Fe}$ and $M_{Tb}$ of our disordered thin films are smaller than typical bulk values (i.e., 2.20 $\mu_B$/Fe or 9.72 $\mu_B$/Tb[8]). The magnetization compensation point ($x_M \approx 47$ for our 20 nm FeTb samples at room temperature), at which $M_s$ vanishes and $H_c$ appears to diverge, also differs from previously reported values for 20 nm FeTb grown on Pt ($x_M \approx 25$)[28-30]. However, it is common that the magnetic properties of thin films of ferrimagnetic alloys can vary depending on growth protocols and substrate choices. For example, $x_M$ of CoTb at room temperature has been reported to be $\approx 22$[13], $\approx 35$[31], $\approx 44$[32] when the CoTb is grown on Ta, Pt and SiN, respectively. We also observe that the magnetic properties of our FeTb samples are sensitive to the layer thickness [Supplementary Fig. S3(c)-(e)], in line with previous reports[30]. As expected, $R_{AHE}$ for the 20 nm

FeTb is negative for $x < x_M$ but positive for $x > x_M$ (see Supplementary Fig. S4(a) for more details) because, when $x < x_M$ ($x > x_M$), the Fe moment is parallel (antiparallel) to the total magnetization and to a strong applied perpendicular magnetic field ($H_z$). The 3$d$ states of Fe govern the anomalous Hall effect because the 4$f$ states of Tb is expected to be located well below the Fermi level and are less involved in transport phenomena of FeTb.

**Giant bulk spin-orbit torque**

We measure the efficiencies of SOTs in the perpendicularly magnetized FeTb samples using the polar-angle dependent harmonic Hall voltage response (HHVR) technique[33,34], after carefully taking into account current-induced heating and thermoelectric effects (Supplementary Sec. 5 and 6). This HHVR technique is accurate when the magnetization rotates coherently at small polar angles ($\theta_M$). This condition is fulfilled in the FeTb samples, as indicated by a well-defined parabolic scaling of the first harmonic Hall signal versus $\theta_M$ (Supplementary Fig. S8). To determine the dampinglike SOT, we rotate the magnetization by scanning a fixed magnitude of magnetic field ($H_{xz}$) relative to the sample at small values of $\theta_M$ in the $x$-$z$ plane (Fig. 2a), and collect the first and the second HHVRs, $V_\omega$ and $V_{2\omega}$, as a function of $\theta_M$ under the excitation of a low-frequency sinusoidal electric field $E$ in the $x$ direction. As we discuss in detail in the Supplementary Sec. 6, the HHVR signals are given by

$$V_\omega = V_{AHE} \cos\theta_M, \quad (1)$$

$$V_{2\omega} \approx (\frac{1}{2}V_{AHE}\frac{H_{DL}}{H_k+H_{xz}} + V_{ANE,z})\sin\theta_M + V_{ANE,x}, \quad (2)$$

where $V_{AHE}$ is the anomalous Hall voltage, $V_{ANE,z(x)}$ is the anomalous Nernst voltage induced by an out-of-plane (in-plane) temperature gradient, and $H_{DL}$ is damping-like effective SOT field. As shown in Fig. 2(b), the measured $V_{2\omega}$ varies linearly with $\sin\theta_M$ for each fixed magnitude of $H_{xz}$. The value of $H_{DL}$ can be obtained from the fits of data to Eq. (2) as shown in Fig. 2(c). In this determination we ignore the so-called "planar Hall correction"[35] because the planar Hall resistance ($R_{PHE}$) samples is negligibly small compared to $R_{AHE}$ ($|R_{PHE}/R_{AHE}| \leq 0.04$, Supplementary Fig. S9), and even if this were not the case the planar Hall correction is generally found to give incorrect values when it is not negligible[36-38]. As shown in Fig. 2(d), $H_{DL}$ for the FeTb single layers with different $x$ increase much more slowly than $1/M_s$ scaling upon approaching the magnetization compensation point ($x_M \approx 47$), which is in sharp contrast to the behavior observed for HM/FM bilayers in which $H_{DL}$ is proportional to $1/M_s$. $H_{DL}$ reverses sign between $x_M$ and the angular momentum compensation point ($x_A \approx 38$, see below).

Using the obtained $H_{DL}$ values, we calculate $\xi_{DL}^j$ of these FeTb single layers following:

$$\xi_{DL}^j \equiv j_s/j = (2e/\hbar)H_{DL}M_s t/j \quad (3)$$

where $e$ is the elementary charge, $\hbar$ the reduced Plank's constant, $M_s$ the saturation magnetization of the spin current detector, $t$ the thickness of the spin current detector, $j_s$ the spin current density absorbed by the spin current detector, and $j = E/\rho_{xx}$ the current density in the spin current generator with electrical resistivity $\rho_{xx}$ (Supplementary Fig. S10). As we justify in the Supplementary Sec. 10, Eq. (3) holds for FIMs regardless of the sign of effective gyromagnetic ratio ($\gamma_{eff}$). As plotted in Fig. 2(e), $\xi_{DL}^j$ of the 20 nm FeTb first increases rapidly from +0.11 at $x = 29$ to +0.41 at $x = 37$, then (like $H_{DL}$) suddenly becomes negative for $38 < x < 47$, and finally becomes positive again and starts to decrease from the value + 0.16 at $x$ = 49 upon further increase of $x$ (see Supplementary Sec. 11 for more details of the torque determination). This sign reversal of the dampinglike spin-orbit torque is reaffirmed by the opposite polarity of the current-induced magnetization switching of the $Fe_{57}Tb_{43}$ and the $Fe_{67}Tb_{33}$ (Supplementary Sec. 16). We find that the sign reversal appears to be correlated to that of the angular momentum. In Fig. 2(f), we show the effective gyromagnetic ratio for the 20 nm FeTb with different composition as calculated using the relation [39-41] $\gamma_{\text{eff}} = (m_{\text{Fe}} - m_{\text{Tb}})/(m_{\text{Fe}}/|\gamma_{\text{Fe}}| - m_{\text{Tb}}/|\gamma_{\text{Tb}}|)$, the magnetic moments of the two sublattices $m_{\text{Fe}} = (1\text{-}0.01x)M_{\text{Fe}}$ and $m_{\text{Tb}} = 0.01xM_{\text{Tb}}$, and the individual gyromagnetic ratios $\gamma_{\text{Fe}}$ = -2.1$\mu_B/\hbar$[42] and $\gamma_{\text{Tb}}$ = -1.5$\mu_B/\hbar$[43]. The composition of the angular momentum compensation point, where the total angular momentum $S = m_{\text{Fe}}/|\gamma_{\text{Fe}}| - m_{\text{Tb}}/|\gamma_{\text{Tb}}|$ is zero, is estimated to be $x_A \approx$ 38.5 for the 20 nm FeTb at the room temperature. We note that $\gamma_{\text{eff}}$, $x_A$, and $x_M$ in Figs. 2(d)-2(f) are only the 20 nm FeTb samples and different from that for the thicker films (e.g. for the films in Fig. 2(g), $x_A < x_M < 43$). The FeTb also shows a field-like torque that is relatively small compared to the damping-like torque (Supplementary Fig. S12).

**Bulk characteristics and microscopic origin**

To analyze these data, we first show that the strong damping-like torque we observe within FeTb is a bulk effect. Qualitatively similar to previous measurements of CoPt single layers[26], $\xi_{\text{DL}}^j$ of the FeTb layers increases linearly with layer thickness when the thickness is greater than about 40 nm as shown in Fig. 2(g) for $Fe_{57}Tb_{43}$, yielding in the bulk limit a SOT efficiency per thickness of $\xi_{\text{DL}}^j/t$ = 0.036 ± 0.008 nm$^{-1}$. This behavior is not consistent with an interfacial torque, for which $\xi_{\text{DL}}^j$ should be approximately independent of the magnetic-layer thickness[24,44]. We also find that this torque is insensitive to the details of the sample interfaces because we measure essentially the same value of $\xi_{\text{DL}}^j$ from symmetric MgO/$Fe_{61}Tb_{39}$ 20 nm/MgO samples and asymmetric SiO$_2$/$Fe_{61}Tb_{39}$ 20 nm/MgO samples. We thus conclude from these characteristics that the damping-like spin torque in FeTb single layers is a bulk effect. This bulk torque is microscopically distinct from the previously reported "interface-engineered" self-torque concluded from a study of GdFeCo [22].

We suggest that the source of the strong damping-like SOT in the perpendicularly magnetized FeTb is most likely a strong conventional bulk spin Hall effect (SHE). We have considered the possibility of origins associated with the anomalous Hall effect or planar Hall effect, but these can only generate spin polarization collinear with the magnetization[45-47]. Magnetic and antiferromagnetic spin Hall effects[48,49] are also not relevant because they are odd under time reversal, while we find the damping-like torque efficiencies generated in FeTb for a given applied electric field does not reverse orientation when the magnetization reverses. We have further verified the existence of a strong SHE in FeTb by measuring the spin current emitted by FeTb layers. We performed thickness-dependent spin-torque ferromagnetic resonance (ST-FMR) experiments[50,51] on a control sample of $Fe_{50}Tb_{50}$ (20 nm)/Ti (1 nm)/Fe (3.8-10.5 nm) and used the in-plane magnetized Fe layer to detect the spin current emitted from the FeTb (Fig. 2(h)). Here, the 1 nm Ti spacer layer was used to suppress the exchange coupling between the FeTb and the Fe layers (Supplementary Fig. S13(b)). The PMA FeTb produces no measurable FMR excitation under the condition of small in-plane magnetic field, so the ST-FMR signal we measure from the FeTb/Ti/Fe trilayers corresponds only to magnetic dynamics from the Fe layer. If we define the apparent FMR spin-torque efficiency ($\xi_{FMR}$) from the ratio of the symmetric and anti-symmetric components of the magnetoresistance response of the ST-FMR (Supplementary Sec. 13), the actual efficiency of the damping-like torque acting on the Fe layer due to the spin current emitted by the Fe$_{50}$Tb$_{50}$ ($\xi_{DL,ext}^j$) can be determined by the method of ref. [51] based on the y-axis intercept in a linear fit of $1/\xi_{FMR}$ versus $1/t_{Fe}$. As shown in Fig. 2(i), we measure $\xi_{DL,ext}^j$ = 0.16 ±0.02 for Fe$_{50}$Tb$_{50}$/Ti/Fe, which is 3 times stronger than that of Pt/Fe bilayers (0.051 ±0.002, also shown in Fig. 2(i)) for Pt with resistivity 38 μΩ cm. We have also measured $\xi_{DL,ext}^j$ for $x$ = 43 (where $\xi_{DL}^j$ is negative) and for $x$ = 29 (on the other side of the angular momentum compensation point where $\xi_{DL}^j$ is positive again), and we find that the sign of $\xi_{DL,ext}^j$ is unambiguously positive at all three concentrations. The value of $\xi_{DL,ext}^j$ that we quote for Fe$_{50}$Tb$_{50}$ only represents a lower bound for the internal value of spin Hall ratio because the torque applied to the Fe is reduced by spin attenuation in the Ti spacer[26,52], interfacial spin backflow[11], and spin memory loss[53]. Spin memory loss, in particular, should be significant at the Ti/Fe interface because it possesses strong interfacial spin-orbit coupling[11] as indicated by the large interfacial magnetic anisotropy energy density of 1.43 ±0.05 erg/cm$^2$ (Supplementary Sec. 14).

A non-zero SOT in a single magnetic layer requires that the sample structure is not symmetric relative to a mirror parallel to the sample plane[16,26]. The required broken symmetry within the FeTb layers seems unrelated to any vertical composition gradient because there is no evidence of a composition gradient in the EELS studies of our films. We also find that a deliberately introduced vertical composition gradient does not enhance the damping-like torque in FeTb. A control sample of 20 nm thick Fe$_{100-x}$Tb$_x$ in which $x$ varied from 27 to 41 with thickness gave $\xi_{DL}^j$ of 0.05±0.01 (Supplementary Fig. S15), which is similar to the averaged value of whole film over the thickness using the composition-dependent values in Fig. 2(e), but significantly smaller in magnitude than -0.46 for 20 nm Fe$_{61}$Tb$_{39}$. In addition, the source of the symmetry breaking is not a vertical thermal gradient because the magnitude of $H_{DL}$ scales in proportion to the applied electric field (Supplementary Fig. S7) and thus $\xi_{DL}^j$ is independent of the applied electric field (symmetry breaking due to Joule heating would give $\xi_{DL}^j \propto E^2$).

**Strong composition dependence and sign change**

We now turn to analyze the dependence on $x$ of the bulk anti-damping spin torque efficiency $\xi_{DL}^j$ for Fe$_{100-x}$Tb$_x$ (Fig. 2(e)). We observe that $|\xi_{DL}^j|$ for the 20 nm Fe$_{100-x}$Tb$_x$ samples shows a broad peak around $x$ = 43, suggesting an enhanced SHE in the intermediate composition range, near and between the two compensation points for magnetization and angular momentum. $|\xi_{DL}^j|$ reaches 0.5 for 20 nm Fe$_{57}$Tb$_{43}$ films and 3 for 90 nm Fe$_{57}$Tb$_{43}$ films. Our result differs from the case of GdFeCo, which was reported to have zero self-torque at the compensation point of angular momentum in a previous temperature-dependence study [22].

The sign change of $\xi_{DL}^j$ that we observe in the 20 nm Fe$_{100-x}$Tb$_x$ samples between the compensation points for magnetization and angular momentum appears to be correlated with the relative orientation of the magnetization and angular momentum vector. Outside the region between the two compensation points the magnetization ($m_{Fe} - m_{Tb}$) and angular momentum ($s_{Fe} - s_{Tb}$) are antiparallel ($\gamma_{eff}$<0), but between the compensation points the total magnetization becomes parallel to the total angular momentum ($\gamma_{\text{eff}}>0$). A change in the sign of $\xi_{\text{DL}}^j$ indicates a change in the sign of the spin angular momentum being transferred to the magnet. However, our ST-FMR measurements on Fe$_{100-x}$Tb$_x$/Ti/Fe indicate no sign change in the polarization of the spin current emitted from FeTb regardless of composition. The microscopic origin of the sign change of $\xi_{\text{DL}}^j$ remains a puzzle and worth study in the future.

**Practical impact and self-torque-driven magnetization switching**

From technological point of view, a strong bulk torque can be advantageous by itself or in combination with interface-applied torques for applications, such as perpendicular magnetic recording and chiral domain wall/skyrmion devices, that require relatively large thickness for high thermal stability. The damping-like SOT efficiency per unit thickness that we measure in the bulk limit for Fe$_{57}$Tb$_{43}$, $\xi_{\text{DL}}^j/t \approx 0.036$ nm$^{-1}$, is much greater than previous reports for other magnetic single layers, e.g ~ 0.0017 nm$^{-1}$ for in-plane NiFe[25], ~ -0.008 nm$^{-1}$ for in-plane CoPt[26], ~ 0.005 nm$^{-1}$ for in-plane FePt[16], and ~ 0.016 nm$^{-1}$ for perpendicular GdFeCo[22]. Here we do not compare our $\xi_{\text{DL}}^j/t$ result with those out-of-plane HHVR results obtained by applying a large "planar Hall correction" (e.g. 0.045 nm$^{-1}$ for $L1_0$-FePt single crystals in ref. [19]), because, as we noted above, the planar Hall correction is generally found to give incorrect values when it is not negligible [36-38].

The bulk SOT of FeTb is sufficiently strong to drive SOT switching of layers with very large thicknesses and strong PMA. In Fig. 3(a)-3(d) we compare magnetic-field-driven switching and SOT switching for both a 20 nm Fe$_{67}$Tb$_{33}$ device ($x < x_A$, Fe-dominated, $M_s$ = 250 emu/cm$^3$, $H_k$=33.5 kOe, $H_c$=1.59 kOe) and a 20 nm Fe$_{42}$Tb$_{58}$ device ($x > x_M$, Tb-dominated, $M_s$= 164 emu/cm$^3$, $H_k$=17.5 kOe, $H_c$=1.72 kOe) as two representative examples. Figs. 3(a) and 3(b) show the Hall resistance ($R_H$) of the samples as a function of $H_z$, which indicate sharp full switching for both samples with $\Delta R_H = 2R_{AHE} = +11$ Ω (-12.2 Ω) for the Fe- (Tb-) dominated sample. In Figs. 3(c) and 3(d), we show $R_H$ of the two samples measured following the application of sequences of current pulses of different amplitudes (0.2 seconds in duration) under the application of a constant symmetry-breaking in-plane bias field $H_x$ along the current direction ($x$ direction). We measure a switching current density of only $(8.2 \pm 1.2) \times 10^6$ A/cm$^2$ for the 20 nm Fe$_{67}$Tb$_{33}$ and $(5.5 \pm 0.2) \times 10^6$ A/cm for the 20 nm Fe$_{39}$Tb$_{61}$. The current-driven switching is only partial (~16% of the full value of $\Delta R_H$ for magnetic-field-driven switching), likely because the non-uniform pinning impedes free motion of domain walls in this domain-wall-mediated switching regime. Full current-driven reversal is likely still possible in nanodot devices with improved magnetic homogeneity as recently demonstrated in CuPt/CoPt bilayers[54]. Here, the switching chirality is opposite for the Fe-dominated Fe$_{67}$Tb$_{33}$ and Tb-dominated Fe$_{42}$Tb$_{58}$, i.e., clockwise (anti-clockwise) for the former but anti-clockwise (clockwise) for the latter when $H_x > 0$ (Fig. 3(c)-3(d)). This is because the SOT fields are of the same sign for the two samples ($\xi_{\text{DL}}^j > 0$), but the anomalous Hall resistances are of opposite signs ($\Delta R_H > 0$ for Fe$_{67}$Tb$_{33}$, but <0 for Fe$_{42}$Tb$_{58}$). We also note that the 20 nm Fe$_{57}$Tb$_{43}$ ($x_A < x < x_M$, $\Delta R_H > 0$, $\xi_{\text{DL}}^j < 0$) can be also switched at a low current density of $(5.5 \pm 0.1) \times 10^6$ A/cm (Supplementary Fig. S16(c)), but the switching polarity is opposite to that of the Fe$_{67}$Tb$_{33}$ ($x < x_A$, $\Delta R_H > 0$, $\xi_{\text{DL}}^j > 0$) due to the negative sign of the bulk spin-orbit torque in the Fe$_{57}$Tb$_{43}$.

## Conclusion

We have demonstrated a giant damping-like SOT arising from the SHE in composition-uniform, amorphous, sputter-deposited ferrimagnetic $Fe_{100-x}Tb_x$ single layers with giant PMA. This bulk torque exhibits no apparent correlation to the interfaces or the absence/presence of a composition gradient. The torque reaches a constant value of efficiency per unit layer thickness in the bulk limit, $\xi_{DL}^j/t \approx 0.036$ nm$^{-1}$. This is more than twice greater any previous report for other magnetic single layers. The torque varies strongly with composition and achieves giant efficiencies $|\xi_{DL}^j|$ of 0.5 for 20 nm $Fe_{61}Tb_{39}$ and 3 for 90 nm $Fe_{57}Tb_{43}$. Interestingly, the torque becomes negative in sign in the intermediate composition range where total angular momentum becomes parallel to the magnetization rather than antiparallel. We also show that the bulk SOT can drive switching in tens of nm thick FeTb layers with strong PMA and high coercivity. For example, the bulk SOT can switch a 20 nm FeTb at very low current densities of a few MA/cm$^2$. Our findings of giant bulk SOT efficiency and intriguing torque-compensation correlation will stimulate study of such unique spin-orbit phenomena in a variety of ferrimagnetic hosts. Our work suggests a promising strategy for self-driven-switching perpendicular ferrimagnetic devices with low power, high density, and straightforward integration with CMOS circuits because there is no requirement for epitaxy or composition gradient.

## Data availability

The data that support this study are available from the corresponding author upon reasonable request.

## Acknowledgements

The authors thank Robert A. Buhrman for support. This work was funded in part by the Office of Naval Research (N00014-19-1-2143), in part by the Defense Advanced Research Projects Agency (USDI D18AC00009), and in part by the NSF MRSEC program (DMR-1719875) through the Cornell Center for Materials Research. Device fabrication was performed at the Cornell Nanofabrication Facility, in part by the NSF (NNCI-2025233) as part of the National Nanotechnology Coordinated Infrastructure, and in part by the Strategic Priority Research Program of the Chinese Academy of Sciences (XDB44000000). Q. Liu acknowledges the financial support by the China Scholarship Council (File No. 201906460052).

## Conflict of Interest

The authors declare no conflict of interest.

## Additional Information

Supplementary Information is available for this paper at xxxxx.

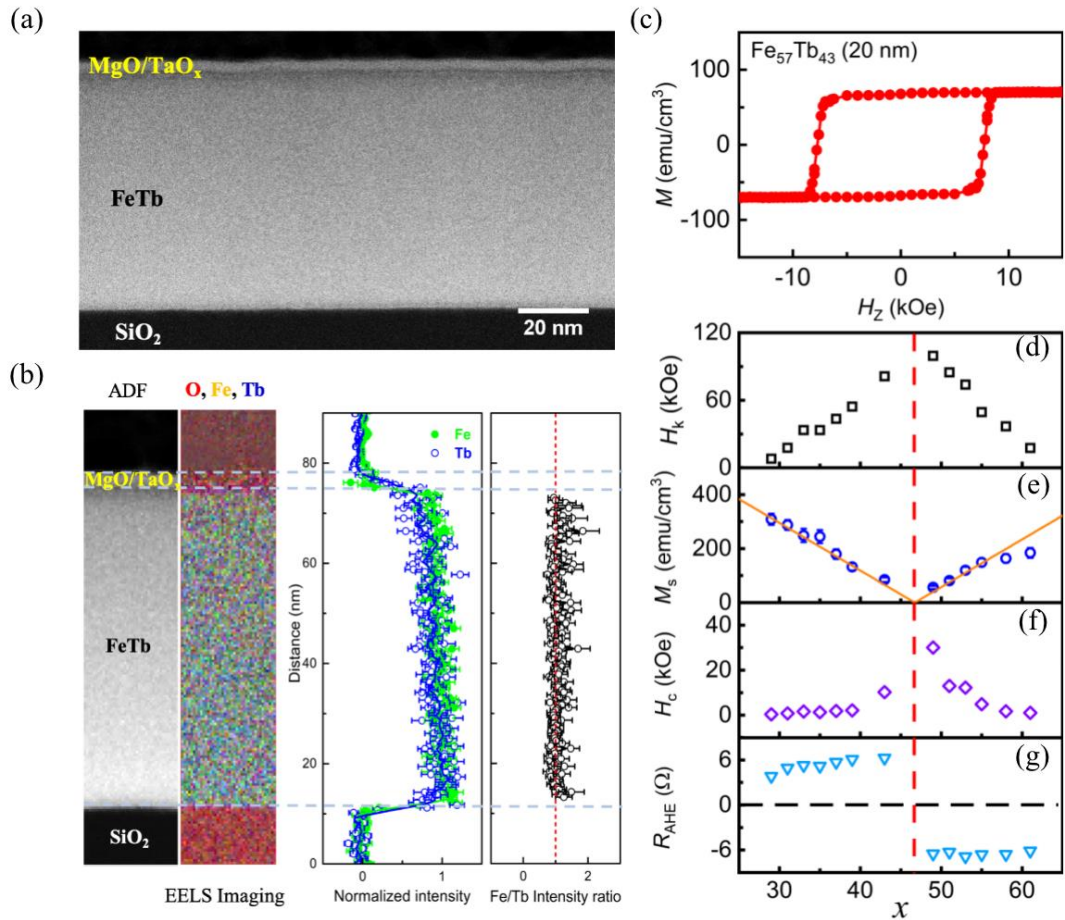

Fig. 1 Sample characterizations. (a) High-angle dark-field cross-sectional STEM image. (b) Depth profile of the EELS intensity for O, Fe, and Tb, showing the absence of any composition gradient in the $Fe_{57}Tb_{43}$ layer. (c) Out-of-plane magnetization curve for the 20 nm $Fe_{57}Tb_{43}$ sample. (d) Perpendicular magnetic anisotropy field ($H_k$), (e) Saturation magnetization ($M_s$), (f) Coercivity ($H_c$), and (g) Anomalous Hall resistance ($R_{AHE}$) of $Fe_{100-x}Tb_x$ films with varying Tb concentration ($x$). Here, the values of $H_k$, $H_c$, and $R_{AHE}$ are determined from transport measurements.

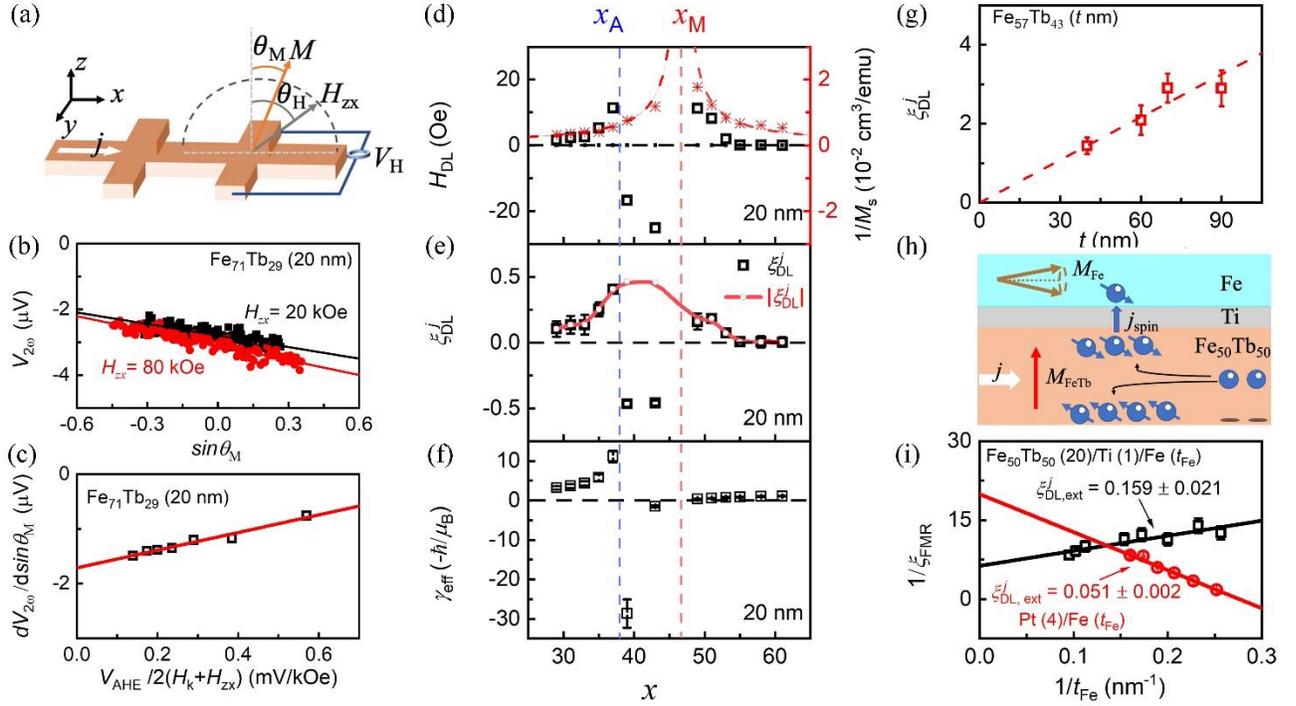

Fig. 2 Spin-orbit torques. (a) Geometry of the HHVR measurement. (b) Second HHVR $V_{2\omega}$ vs $\sin\theta_M$ for a 20 nm $Fe_{71}Tb_{29}$ sample under for constant magnitudes of applied magnetic field $H_{zx} = 20$ and 80 kOe. (c) $dV_{2\omega}/d\sin\theta_M$ vs $V_{AHE}/2(H_k+H_{zx})$ for 20 nm $Fe_{71}Tb_{29}$. Dependence on the Tb concentration $x$ for (d) the damping-like effective SOT field $H_{DL}$, (e) the damping-like torque efficiencies per current density $\xi^j_{DL}$ and $|\xi^j_{DL}|$, and (f) the calculated value of $\gamma_{eff}$ for the 20 nm $Fe_{100-x}Tb_x$. In (d)-(f), the dashed lines indicate the angular momentum compensation point $x_A$ (blue dashed line) and the magnetization compensation point $x_M$ (red dashed line). (g) $\xi^j_{DL}$ vs. the thickness ($t$) of $Fe_{47}Tb_{43}$ samples ($\xi^j_{DL} > 0$ because the composition $x = 43$ is located in the Tb-dominated regime when the thickness is greater than ≈30 nm, see the Supplementary Fig. S4(e)). (h) Schematic of ST-FMR measurements on $Fe_{50}Tb_{50}$ (20 nm)/Ti (1 nm)/Fe ($t_{Fe}$) samples. (i) Inverse FMR efficiency ($1/\xi_{FMR}$) vs. inverse Fe thickness ($1/t_{Fe}$) for the $Fe_{50}Tb_{50}$ (20 nm)/Ti (1 nm)/Fe ($t_{Fe}$ nm) sample and a control sample Pt (4 nm)/Fe ($t_{Fe}$).

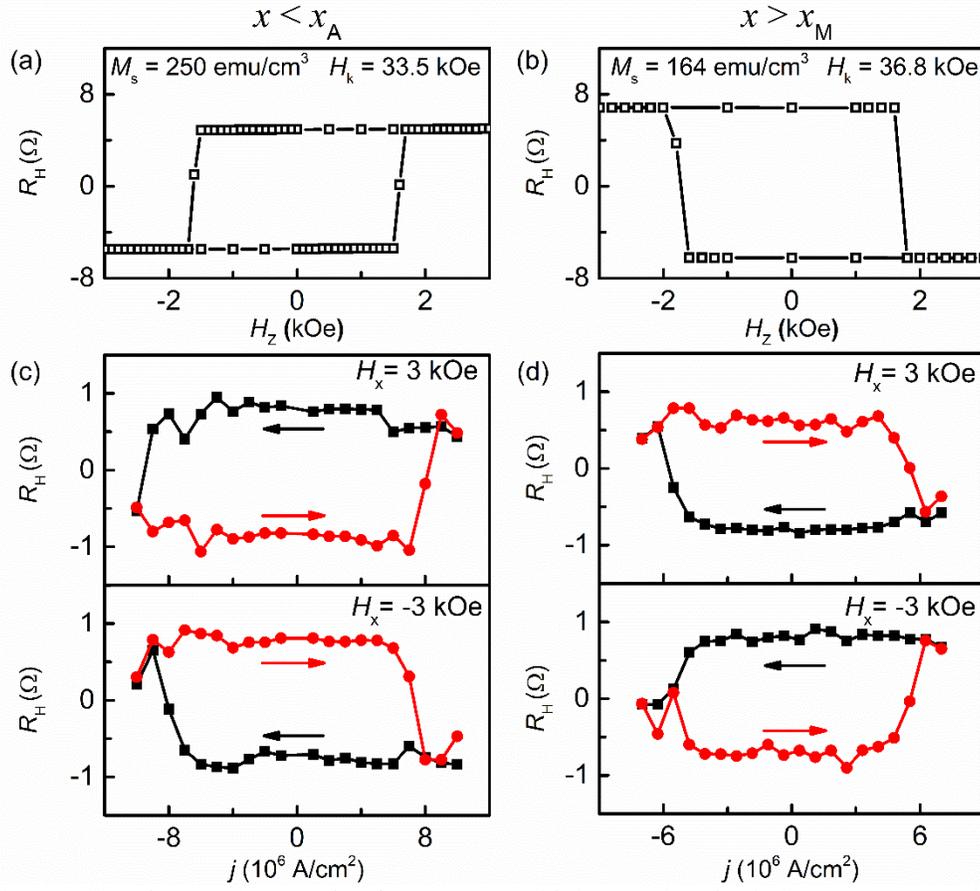

Fig. 3. Anomalous Hall resistance hysteresis of (a) a 20 nm thick $Fe_{67}Tb_{33}$ single layer ($x<x_A$, Fe-dominated, $M_s =$ 250 emu/cm$^3$, $H_k$=33.5 kOe, $H_c$=1.59 kOe) and (b) a 20 nm thick $Fe_{42}Tb_{58}$ single layer ($x>x_M$, Tb-dominated, $M_s=$ 164 emu/cm$^3$, $H_k$=17.5 kOe, $H_c$=1.72 kOe). Current induced magnetization switching of (c) the $Fe_{67}Tb_{33}$ and (d) the $Fe_{42}Tb_{58}$, under a constant in-plane bias field $H_x = \pm 3$ kOe that overcomes the DMI field within the domain walls.

Supplementary Information for

# Giant bulk spin-orbit torque and efficient electrical switching in single ferrimagnetic FeTb layers with strong perpendicular magnetic anisotropy

Qianbiao Liu[1,2+], Lijun Zhu[1,2+*], Xiyue S. Zhang[2], David A. Muller[2], Daniel C. Ralph[2,3]

1. *State Key Laboratory of Superlattices and Microstructures, Institute of Semiconductors, Chinese Academy of Sciences, Beijing 100083, China*
2. *Cornell University, Ithaca, New York 14850, USA*
3. *Kavli Institute at Cornell, Ithaca, New York 14850, USA*

[+]These authors contributed equally to this work.
*ljzhu@semi.ac.cn

**Table of Contents:**

**Section 1. Sample fabrication and characterizations**

A series of $Fe_{100-x}Tb_x$ (FeTb for short) single layers with different fixed Tb volume percentages ($x$ = 15-78) were deposited on oxidized Si substrates by co-sputtering at room temperature, and then protected by capping with MgO (1.6 nm)/Ta (1.6 nm) without breaking vacuum. The Tb volume percentages were calibrated from the fraction of the Tb deposition rates over the total deposition rate during the co-sputtering deposition. The argon pressure was 2 mTorr during the sputtering process, and the base pressure was ~$10^{-9}$ Torr. To make devices for determining the efficiencies of the spin-orbit torques by harmonic Hall voltage response (HHVR) measurements and spin-torque ferromagnetic resonance (ST-FMR) measurements, the layers were patterned by photolithography and ion milling into Hall bars (5×60 μm$^2$, Fig. S1a) and simple microstrips (10×20 μm$^2$, Fig. S1b) followed by deposition of 5 nm Ti and 150 nm Pt as electrical contacts.

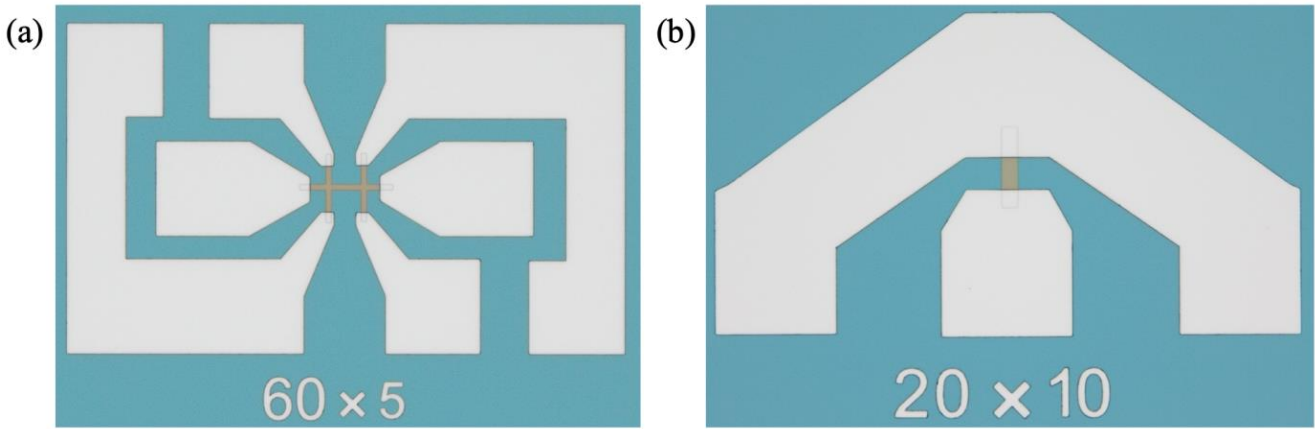

**Figure S1.** Optical images of (a) a Hall bar device and (b) a ST-FMR device.

The magnetic properties of the films were characterized using a superconducting quantum interference device-vibrating sample magnetometer (SQUID-VSM). The sample for scanning transmission electron microscopy (STEM) imaging and electron energy loss spectrum (EELS) measurements was thinned using a focused ion beam (FIB) system using a FEI/Thermo Fisher Titan Themis STEM system at 300 kV. Anomalous Hall resistance and planar Hall resistance were measured using a physical property measurement system (PPMS). During HHVR measurements, a Signal Recovery DSP Lock-in Amplifier Model 7625 was used to source a sinusoidal current onto the Hall bars and to detect the first and second harmonic Hall voltage responses. During the switching measurements, a pulsed write current with a duration of 0.2 ms was sourced to the Hall bar devices using a Keithley 2400. The anomalous Hall voltage was detected either by a Keithley 2182A after each pulsed write current (with a reading current of 0.1 mA) or by a Signal Recovery DSP Lock-in Amplifier Model 7625 (with a reading excitation of 0.1 V). For ST-FMR measurements, the rf current was sourced by signal generator (E8257D) and the mixing voltage was detected using a Signal Recovery DSP Lock-in Amplifier Model 7625. All the measurements were performed at room temperature.

**Section 2. Thickness gradient of the FIB-thinned STEM sample**

Our standard FIB thinning process typically results in a thickness gradient in the cross-sectional view of STEM samples, with the top side of film thinner than the substrate side. This thickness gradient may cause significant variation of thickness-sensitive signals (e.g. the absolute intensity EELS), but it usually does not affect the relative strengths of the signals from different elements. Therefore, the quasi-linear variation of the absolute EELS intensity of the 66 nm FeTb sample in Fig. 1(b) of the main text is simply due to the thickness gradient formed during the FIB thinning process and should not be misinterpreted as a composition gradient.

For a detailed understanding of the formation of the thickness gradient, we now briefly describe the standard FIB thinning process we employ in this work. We first protect the initial thin-film sample surface by depositing 20-30 nm carbon and 0.8-1 μm Pt and prepare a lamella on the specimen using a Ga ion beam. The lamella is transferred to a TEM grid using a needle and then fixed by sputtering the "paste" Pt (Fig. S2(a)), followed by repeatedly cleaning the cross-section by 30 keV Ga ion milling from both sides until the thickness is down to 200-500 nm. The lamella is then thinned further from both sides using 5 keV Ga ions and smaller milling box (see Fig. S2(b) until the protective Pt layer on the top is reduced to 0-20 nm). Because the lamella is tilted 2-3 degrees during the Ga ion beam thinning process, the end result includes a thickness gradient with the top of the lamella the thinnest.

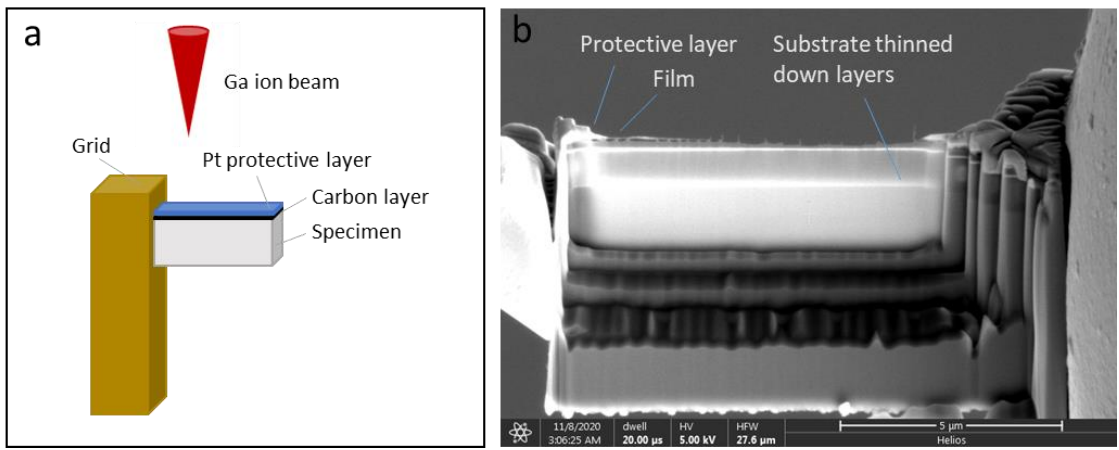

**Figure S2** FIB thinning process. (a) Carton of a lamella attached to the TEM grid before thinning. (b) Scanning electron microscopy of a lamella after thinning from both sides until the top protective Pt layer reaches 0-20 nm. The thinning box is moved gradually toward the top and the inside of the lamella. A thickness gradient with the top side thinnest is finally formed because of the tilted thinning from both sides.

## Section 3. Magnetic properties of Fe$_{100-x}$Tb$_x$ single layers

We obtained FeTb single layers with square perpendicular magnetization hysteresis loops over a wide composition range (the Tb composition $x$ = 29-61, Fig. S3(a)) and over a wide thickness range (even down to 7 nm, Fig. S3(b)). The saturation magnetization and the perpendicular magnetic anisotropy field of the FeTb can be tuned strongly by the composition (Figs. S3(c) and S3(d)) or by the layer thickness (Figs. S3(e)). The thickness dependence of the magnetic properties was interesting and was attributed to a varying alignment of the Tb magnetic moments with the layer thickness in a previous report [S1].

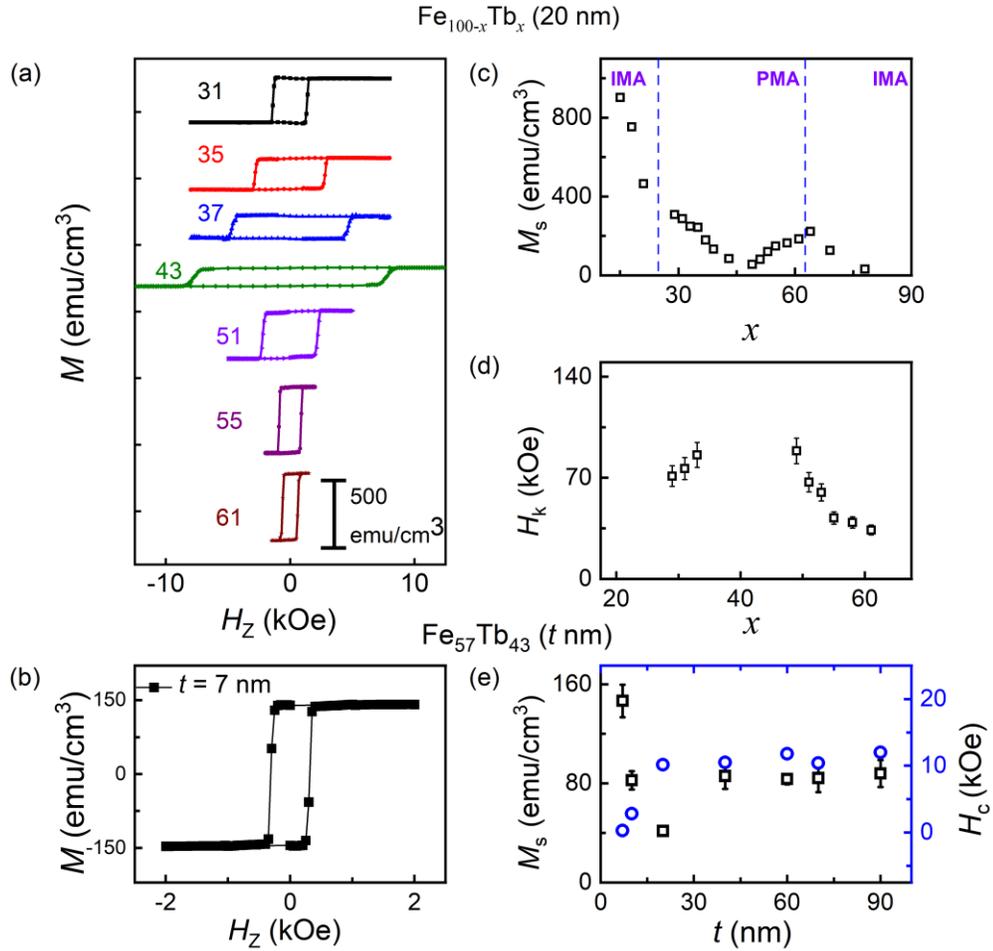

**Figure S3.** Magnetic properties of Fe$_{100-x}$Tb$_x$ single layers as measured by SQUID-VSM. The out-of-plane magnetization hysteresis loops (*M-H* curves) of (a) 20 nm-thick Fe$_{100-x}$Tb$_x$ single layers with different Tb concentrations ($x$ = 31-61) and (b) a 7 nm thick Fe$_{57}$Tb$_{43}$. (c) The saturation magnetization ($M_s$) and (d) the perpendicular magnetic field ($H_k$) of the 20 nm Fe$_{100-x}$Tb$_x$ thin films plotted as a function of the Tb concentration. The values of $H_k$ around $x$ = 47 are not measureable by SQUID. (e) $M_s$ and the perpendicular coercivity of the Fe$_{57}$Tb$_{43}$ films plotted as a function of the layer thickness.

## Section 4. Anomalous Hall resistance and coercivity of Fe$_{100-x}$Tb$_x$ single layers

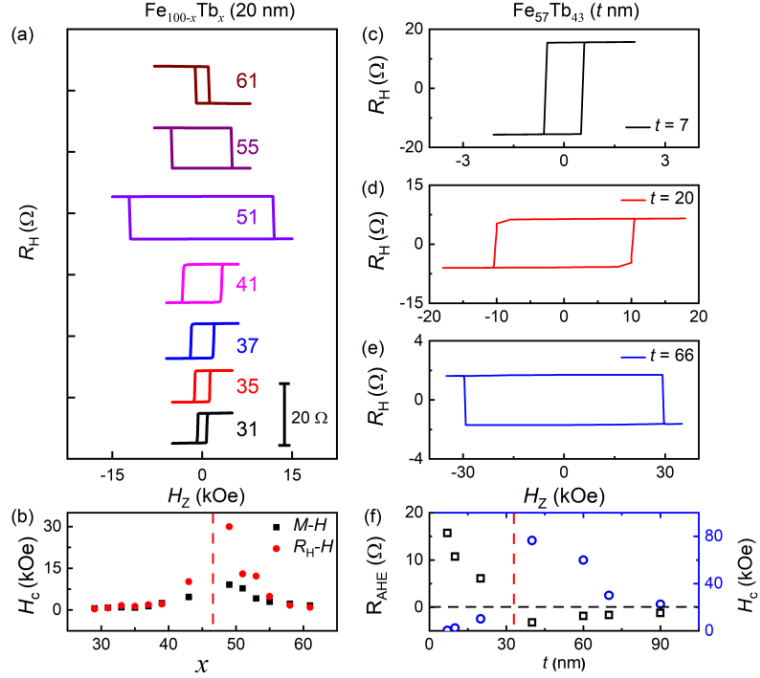

**Figure S4.** (a) Anomalous Hall resistance ($R_H$) hysteresis of 20 nm Fe$_{100-x}$Tb$_x$ with different compositions. $R_H$ is negative for $x < x_M$ and positive for $x > x_M$ ($x_M = 47$ for the 20 nm films). (b) The coercivity of the 20 nm Fe$_{100-x}$Tb$_x$ as measured from the electrical measurement (red points) and magnetic hysteresis measurement (black points). The coercivity difference around the magnetic compensation point is attributed to enhanced domain pinning during device fabrication. Anomalous Hall resistance hysteresis of (c) 7 nm, (d) 20 nm and (e) 66 nm Fe$_{57}$Tb$_{43}$. (f) Anomalous Hall resistance and coercivity of the Fe$_{57}$Tb$_{43}$ films with different thicknesses. Thickness-induced sign reversal of the anomalous Hall resistance indicates a compensation thickness of ≈30 nm for the Fe$_{57}$Tb$_{43}$. The bias current is 0.1 mA during the measurement so that Joule heating is negligible.

## Section 5. Estimation of current-induced temperature increase

We can estimate the temperature increase of our FeTb samples induced by Joule heating during measurements of spin-orbit torque and current-driven switching by simultaneously measuring the overall change in the sample resistance. We first measure the resistance of the Hall bar devices as a function of temperature using a small dc bias of 0.1 mA (see Fig. S5(a) for example of 20 nm Fe$_{59}$Tb$_{41}$) so that the result is not affected by current-induced heating. Then, the temperature rise during a harmonic Hall voltage response measurement can be calibrated according to the resistance. As summarized in Fig. 5(b), the current-induced temperature increase ($\Delta T$) relative to 300 K varies from 11 K to 22 K, depending on the Tb concentration (thus resistivity) of the FeTb. This temperature increase is sufficient to affect the saturation magnetization of FeTb during harmonic Hall voltage response measurements (Fig. S5(c)). Therefore, in the calculation of spin orbit torque efficiency in our samples we have used the magnetization at the calibrated temperature rather than at 300 K.

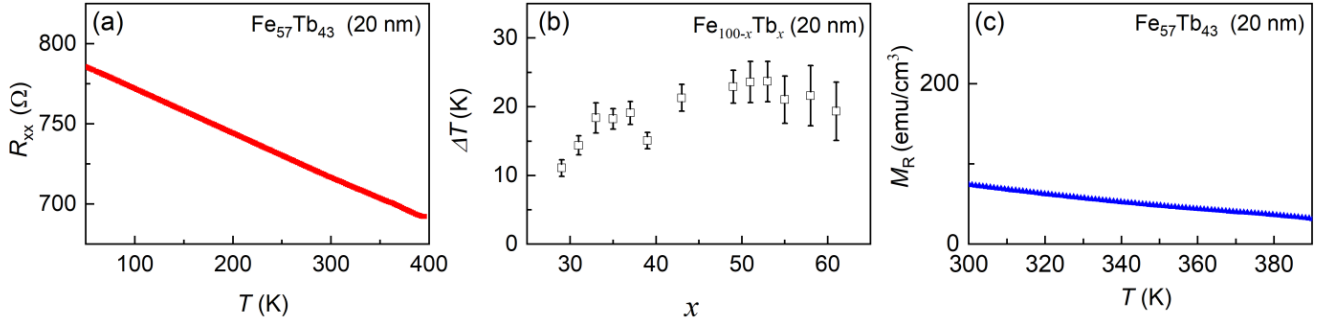

**Figure S5.** Current-induced temperature increase. (a) Temperature dependence of the longitudinal resistance ($R_{xx}$) of a 20 nm thick $Fe_{59}Tb_{41}$ sample measured with a small DC current (0.1 mA in a Hall bar that is 5 μm wide and 60 nm long). (b) The calibrated temperature increase ($\Delta T$) relative to 300 K of the FeTb with different Tb concentration during the harmonic Hall voltage response measurements. (c) Variation of the magnetization of a 20 nm thick $Fe_{59}Tb_{41}$ layer as a function of temperature.

**Section 6. Subtraction of effects of anomalous Nernst voltage from harmonic Hall voltage response**

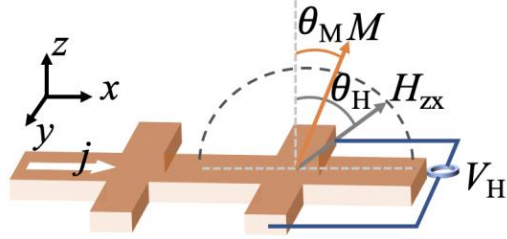

**Figure S6.** Schematic of measurement coordinates.

A charge current flow in a resistive perpendicularly-magnetized Hall bar device can induced thermal gradients in both the out-of-plane ($\nabla T_z$) and in-plane directions ($\nabla T_x$) [S2]. Sizable anomalous Nernst voltages can arise from these thermal gradients and contribute to measurements of the harmonic Hall voltage responses (HHVRs). In a general case, the first and the second HHVRs, $V_\omega$ and $V_{2\omega}$, are given by [S2]:

$$V_\omega = V_{AHE}\cos\theta_M + V_{PHE}\sin^2\theta_M \sin 2\varphi, \qquad (S1)$$

$$V_{2\omega} = V_{2\omega,SOT} + V_{ANE,x}\cos\theta_M + V_{ANE,z}\sin\theta_M\cos\varphi. \qquad (S2)$$

where $\theta_M$ is the polar angle of the magnetization (with $\theta_M = \arccos(V_\omega/V_{AHE})$ if the angle of applied magnetic field is chosen so that $\sin 2\varphi = 0$), $\varphi$ is the azimuthal angle of the magnetization and is the same as that of the magnetic field for a uniaxial anisotropy system, $V_{2\omega,SOT}$ is the second HHVR due to the spin-orbit torques, $V_{AHE}$ is the anomalous Hall voltage coefficient, $V_{PHE}$ is the planar Hall voltage coefficient, $V_{ANE,x} = c_{ANE}\nabla T_x$ and $V_{ANE,z} = c_{ANE}\nabla T_z$ are the anomalous Nernst voltages associated with the in-plane and perpendicular thermal gradients, and $c_{ANE}$ is the anomalous Nernst coefficient. $V_{AHE}$ can be determined from the dependence of $V_\omega$ on a swept out-of-plane magnetic field ($H_z$) or on a small sweep in-plane field $H_x$ if $\theta_M$ is determined. $V_{ANE,z}$ is equal to the value of $V_{2\omega}$ when $\theta_M = 90°$ and $\varphi = 0°$ (current direction), while $V_{ANE,x}$ is equal to the value of $V_{2\omega}$ when $\theta_M = \theta_H = 0°$ (film normal direction). In Eq. (S2) we have ignored possible contributions from the ordinary Nernst effect that tends to be much smaller than the anomalous Nernst effect in samples with metallic ferromagnets [S3].

(i) When a magnetic field applied in the $xz$ plane ($H_{xz}$) tilts the magnetization by a small $\theta_M$, Eqs. (S1) and (S2) can be simplified as:

$$V_\omega = V_{AHE} \cos\theta_M, \tag{S3}$$

$$V_{2\omega} = (\frac{1}{2}\frac{V_{AHE}}{H_k+H_{xz}} H_{DL} + V_{ANE,z}) \sin\theta_M + V_{ANE,x}. \tag{S4}$$

In this case, $H_{DL}$ can be determined from the dependence of $\partial V_{2\omega}/\partial \sin\theta_M$ on the magnitude of $H_{xz}$.

(ii) When a small in-plane field $H_x$ is applied along the current direction (with $H_x/H_k <<1$ so that $\sin\theta_M \approx H_x/H_k$), Eqs. (S1) and (S2) can be simplified as:

$$V_\omega = V_{AHE}(1-\frac{H_x^2}{2H_k^2}), \tag{S5}$$

$$V_{2\omega} = \frac{1}{2}\frac{V_{AHE}}{H_k^2} H_{DL} H_x + \frac{V_{ANE,z}}{H_k} H_x + V_{ANE,x}. \tag{S6}$$

Now, since both the first and the second terms in Eq. (S6) are proportional to $H_x$, the anomalous Nernst contribution to the second HHVR cannot be separated from the dampinglike torque contribution simply from the dependences of $V_{2\omega}$ on $H_x$. Generally, one can define an apparent longitudinal effective field

$$H_L = -2\frac{\partial V_{2\omega}}{\partial H_x}/\frac{\partial^2 V_\omega}{\partial H_x^2} = H_{DL} + 2H_k \frac{V_{ANE,z}}{V_{AHE}}. \tag{S7}$$

Both $H_{DL}$ and the second term of Eq. (S7) are proportional to the magnitude of the applied ac electric field $E$. This suggests that the anomalous Nernst effect should be carefully taken into account in the analysis of resistive magnetic systems with a small SOT field and a large value of $H_k$. This can be done by measuring each of the parameters contributing to the second term in Eq. (S7) separately, as described above. We find that the anomalous Nernst term is significant for some FeTb single layers ($x\leq 51$). As an example, we show the results of $V_{ANE,z}$, $V_{AHE}$, $H_L$, and $H_{DL}$ as a function of $E$ for a 20 nm thick $Fe_{71}Tb_{29}$ in Fig. S7 (a)-(c). However, we find that for the Pt 5 nm/FeTb 20 nm, $H_L$ is the same as $H_{DL}$ within the experimental uncertainty, indicating minimal influence of thermoelectric effect.

(iii) When a small in-plane field $H_y$ is applied transverse to the current ($H_y/H_k <<1$ so that $\sin\theta_M \approx H_y/H_k$), Eq. (S2) can be simplified as:

$$V_\omega = V_{AHE}(1-\frac{H_y^2}{2H_k^2}). \tag{S8}$$

$$V_{2\omega} = \frac{1}{2}\frac{V_{AHE}}{H_k^2} H_{FL} H_y + V_{ANE,x}. \tag{S9}$$

From this equation, the field-like spin-orbit torque field $H_{FL}$ can be determined from the slope of the linear fit of $V_{2\omega}$ vs $H_y$.

$$H_T = -2\frac{\partial V_{2\omega}}{\partial H_y}/\frac{\partial^2 V_\omega}{\partial H_y^2} = H_{FL}. \tag{S10}$$

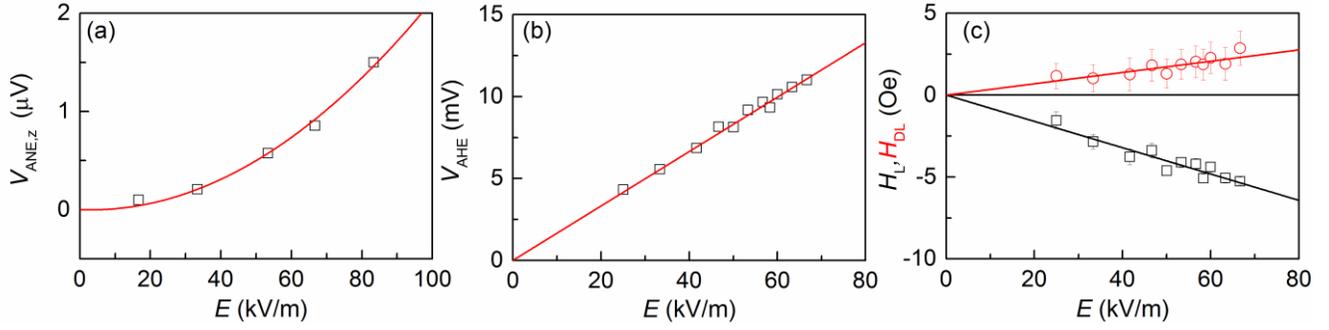

**Figure S7.** Electric field ($E$) dependence of (a) $V_{ANE,z}$, (b) $V_{AHE}$, (c) $H_L$ and $H_{DL}$ for a 20 nm $Fe_{71}Tb_{29}$ layer.

## Section 7. Coherent rotation at small polar angles

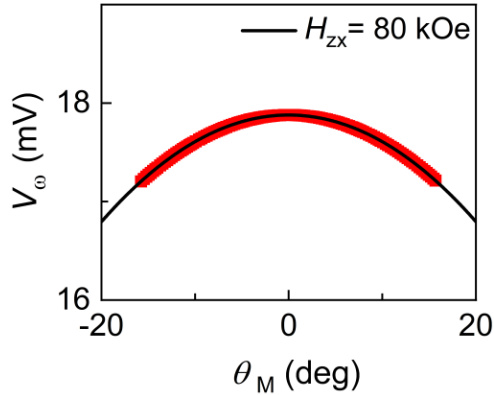

**Figure S8.** Polar-angle dependence of the first harmonic Hall voltage response ($V_\omega$) of a 20 nm $Fe_{71}Tb_{29}$ Hall bar device under an external magnetic field with a constant magnitude $H_{zx}= 80$ kOe in the x-z plane. The solid parabolic line represents the best fit to the expression $V_\omega = V_{AHE}\cos\theta_M \approx V_{AHE}(1-\theta_M^2/2)$, indicating a coherent magnetization rotation at small polar angles.

## Section 8. Planar Hall resistance

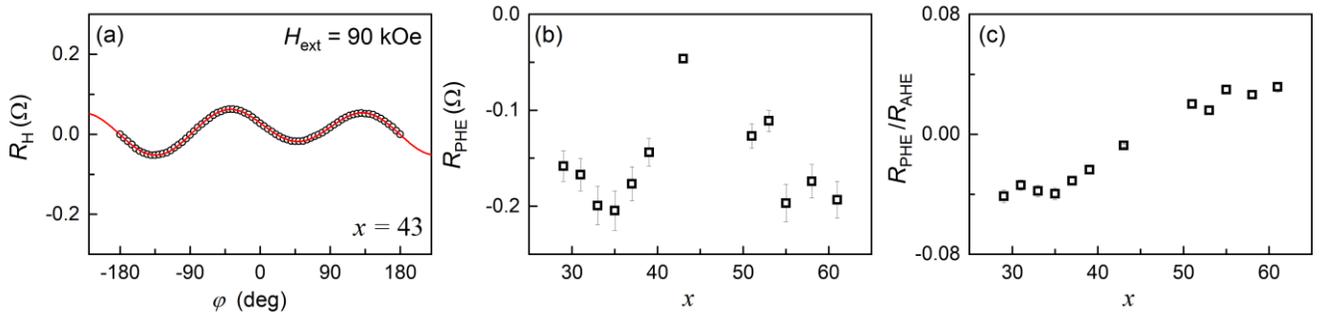

**Figure S9. Planar Hall resistance of a 20 nm thick $Fe_{55}Tb_{45}$ layer.** (a) Hall resistance ($R_H$) for a $Fe_{55}Tb_{45}$ single layer plotted as a function of the in-plane angle ($\varphi$) of an in-plane magnetic field (90 kOe) relative to the current direction. The planar Hall resistance ($R_{PHE}$) is determined from the best fit (red line) of the data to the relation $R_H = R_{PHE}\sin 2\varphi + C\sin\varphi$ (the $C\sin\varphi$ term comes from the small misalignment of the magnetic field with respect to the sample plane). (b) $R_{PHE}$ and (c) the ratio of the planar to anomalous Hall resistances ($R_{PHE}/R_{AHE}$) plotted as a function of the Tb concentration. The $|R_{PHE}/R_{AHE}|$ ratios are $\leq 0.04$ for $Fe_{100-x}Tb_x$.

## Section 9. Resistivities of 20 nm thick $Fe_{100-x}Tb_x$ layers

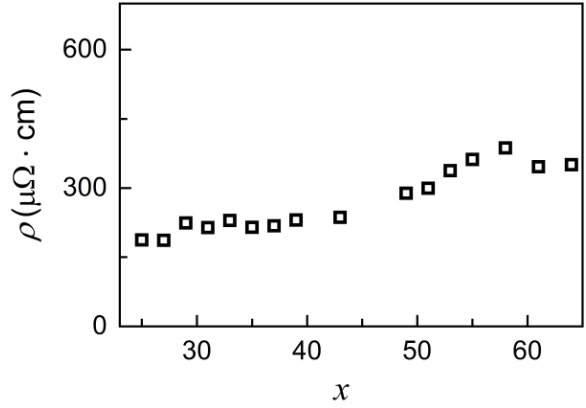

**Figure S10.** Resistivity of the 20 nm thick $Fe_{100-x}Tb_x$ layers with different Tb concentrations.

## Section 10. Validation of Eq. (3) in the main text

We compare the effect of an external magnetic field to the effect a dampinglike spin-orbit torque on a ferrimagnet composed of two oppositely oriented magnetic sublattices (1 and 2) with different gyromagnetic ratios $\gamma_i$ ($i$=1,2). We assume the individual gyromagnetic ratios are negative, so that the angular momentum ($\vec{S}_i$) and magnetization ($\vec{M}_i$) vectors are related as $\vec{S}_i = -\vec{M}_i/|\gamma_i|$ with magnitudes $S_i = M_i/|\gamma_i|$. All calculations below will assume a unit area of a sample thin film.

The LLG equation for an individual sublattice subject only to a magnetic field has the form

$$\frac{d}{dt}\vec{M}_i = -|\gamma_i|M_i\hat{m}_i \times \vec{H} + \alpha_i M_i \hat{m}_i \times \frac{d\hat{m}_i}{dt}. \qquad (S11)$$

where $\alpha_i$ is the Gilbert damping and $M_i$ is the magnetic moment. It will be more convenient to write the combined equation of motion for the two sublattices in terms of the total angular momentum rather than the total magnetization, because the exchange interaction between the two sublattices will conserve total angular momentum, while it does not conserve the total magnetization.

The LLG equation for sublattice 1 can be re-written

$$\frac{d}{dt}\vec{S}_1 = -M_1\hat{s}_1 \times \vec{H} - \frac{\alpha_1}{|\gamma_1|}M_1\hat{s}_1 \times \frac{d\hat{s}_1}{dt}. \qquad (S12)$$

and for sublattice 2

$$\frac{d}{dt}\vec{S}_2 = -M_2\hat{s}_2 \times \vec{H} - \frac{\alpha_2}{|\gamma_2|}M_2\hat{s}_2 \times \frac{d\hat{s}_2}{dt} = +M_2\hat{s}_1 \times \vec{B} - \frac{\alpha_2}{|\gamma_2|}M_2\hat{s}_1 \times \frac{d\hat{s}_1}{dt}. \qquad (S13)$$

Adding these equations, the equation of motion for the total angular momentum of the ferrimagnet subject only to a magnetic field is

$$\frac{d}{dt}(\vec{S}_{\text{total}}) = -(M_1 - M_2)\hat{s}_1 \times \vec{H} - \left(\frac{\alpha_1}{|\gamma_1|}M_1 + \frac{\alpha_2}{|\gamma_2|}M_2\right)\hat{s}_1 \times \frac{d\hat{s}_1}{dt}. \qquad (S14)$$

If the ferrimagnet is also subject to an antidamping spin-orbit torque, with spin angular momentum in the $\hat{\sigma}$ direction, we can add this to the equation of motion

$$\frac{d}{dt}(\vec{S}_{\text{total}}) = -(M_1 - M_2)\hat{s}_1 \times \vec{H} - \left(\frac{\alpha_1}{|\gamma_1|}M_1 + \frac{\alpha_2}{|\gamma_2|}M_2\right)\hat{s}_1 \times \frac{d\hat{s}_1}{dt} + \frac{\hbar}{2e}\xi^j_{\text{DL}}j_e(\hat{s}_1 \times \hat{\sigma} \times \hat{s}_1). \qquad (S15)$$

Since $\vec{S}_{\text{total}} = (S_1 - S_2)\hat{s}_1 = \left(\frac{M_1}{|\gamma_1|} - \frac{M_2}{|\gamma_2|}\right)\hat{s}_1 = (M_1 - M_2)\hat{s}_1/\gamma_{\text{eff}}$, where we define $\gamma_{\text{eff}} \equiv (M_1 - M_2)/\left(\frac{M_1}{|\gamma_1|} - \frac{M_2}{|\gamma_2|}\right)$, in terms of unit vectors Eq. (S15) can be rewritten as

$$\frac{d}{dt}(\hat{s}_1) = -\gamma_{\text{eff}}\, \hat{s}_1 \times \vec{H} - \alpha_{\text{eff}}\hat{s}_1 \times \frac{d\hat{s}_1}{dt} + \gamma_{\text{eff}} \frac{\hbar}{2e}\left(\frac{1}{M_1-M_2}\right)\xi_{\text{DL}}^{j} j_e(\hat{s}_1 \times \hat{\sigma} \times \hat{s}_1). \tag{S16}$$

Here $\alpha_{\text{eff}} = \gamma_{\text{eff}}\left(\frac{\alpha_1}{|\gamma_1|}M_1 + \frac{\alpha_2}{|\gamma_2|}M_2\right)/(M_1 - M_2)$.

From Eq. (S16), we can read off that the spin-transfer torque is equivalent to an effective magnetic field

$$\vec{H}_{\text{DL}} = -\frac{\hbar}{2e}\left(\frac{1}{M_1-M_2}\right)\xi_{\text{DL}}^{j} j_e(\hat{\sigma} \times \hat{s}_1) = \frac{\hbar}{2e}\frac{1}{|M_1-M_2|}\xi_{\text{DL}}^{j} j_e(\hat{\sigma} \times \hat{m}). \tag{S17}$$

If the effective magnetic field is measured relative to the direction $\hat{\sigma} \times \hat{m}$, we therefore have that

$$\xi_{\text{DL}}^{j} = \frac{2e}{\hbar}\frac{H_{DL}|M_1-M_2|}{j_e}, \tag{S18}$$

With the identification that the total magnetization per unit area $|M_1 - M_2|$ is equal to $M_s t$, the measured magnetization per unit area, one obtain

$$\xi_{\text{DL}}^{j} = \frac{2e}{\hbar}\frac{H_{\text{DL}} M_s t}{j_e}. \tag{S19}$$

This is the same as Eq. (3) in the main text, and indicates that $\xi_{\text{DL}}^{j}$ has no dependence on the sign of $\gamma_{\text{eff}}$.

## Section 11. Raw data for representative samples exhibiting different torque signs

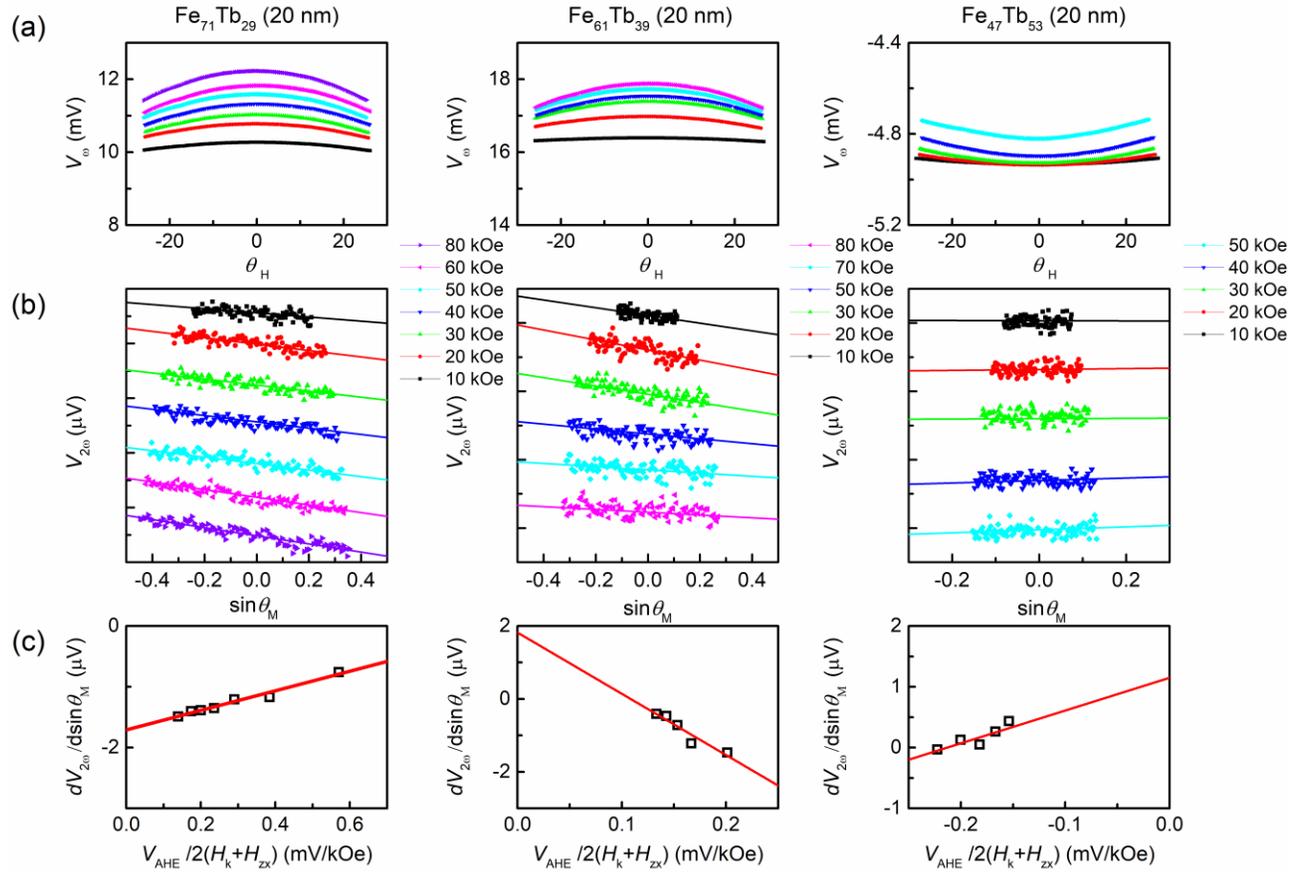

**Figure S11. Raw data for three representative samples:** 20 nm $Fe_{71}Tb_{29}$ ($V_{\text{AHE}} > 0$, $\xi_{\text{DL}}^{j} > 0$), 20 nm $Fe_{61}Tb_{39}$ ($V_{\text{AHE}} > 0$, $\xi_{\text{DL}}^{j} < 0$), and 20 nm $Fe_{47}Tb_{53}$ ($V_{\text{AHE}} < 0$, $\xi_{\text{DL}}^{j} > 0$). (a) Frist ($V_{1\omega}$) vs $\sin\theta_M$, (b) Second HHVR ($V_{2\omega}$) vs $\sin\theta_M$, and (c) $dV_{2\omega}/d\sin\theta_M$ vs $V_{\text{AHE}}/2(H_k+H_{zx})$ for under different applied magnetic field $H_{zx}$. The definition of symbols are the same as those in the Supplementary Section 6.

## Section 12. Field-like spin-orbit torque in $Fe_{100-x}Tb_x$ single layers

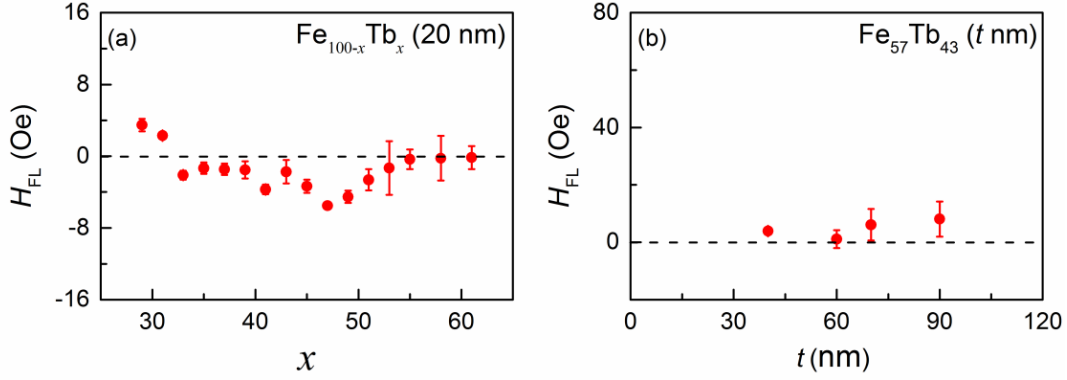

**Figure S12.** (a) Field-like effective torque field for 20 nm thick $Fe_{100-x}Tb_x$ layers with different Tb concentrations. (b) Field-like effective torque field for $Fe_{57}Tb_{43}$ layers with different thicknesses. The field-like torque is relatively small compared to the damping-like torque (Fig. 2(d) of the main text).

## Section 13. Spin-torque ferromagnetic resonance measurements

To confirm the spin Hall effect of the FeTb, we performed spin-torque ferromagnetic resonance (ST-FMR) to measure emission of spin current from FeTb layers that is absorbed by an Fe detector layer. During the ST-FMR measurements, an in-plane magnetic field ($H$) was swept at a fixed angle of 45 °with respect to the magnetic microstrip. As shown in Fig. S13(a), the amplitudes of the symmetric and anti-symmetric components of a ST-FMR spectrum, $S$ and $A$, can be determined by fitting the data to [S4]:

$$V_{\text{mix}} = S \frac{\Delta H^2}{\Delta H^2 + (H - H_r)^2} + A \frac{\Delta H (H - H_r)}{\Delta H^2 + (H - H_r)^2}, \quad (S11)$$

where $\Delta H$ is the FMR linewidth and $H_r$ the resonance field. From the $S/A$ ratio, we can define as an intermediate parameter an effective spin-orbit torque efficiency [S5, S6]:

$$\xi_{\text{FMR}} = \frac{S}{A} \frac{eM_s td}{\hbar} \sqrt{1 + \frac{4\pi M_{\text{eff}}}{H_r}}, \quad (S12)$$

where $e$ is the electron charge, $\hbar$ is the reduced Planck constant, $\mu_0$ is the vacuum permeability, $M_s$ is the saturation magnetization. $t$ is the layer thickness of the magnetic detector, $d$ is the layer thickness of the spin-current-generating layer, and $4\pi M_{\text{eff}}$ is the effective demagnetization field of the magnetic detector. The true damping-like spin-torque efficiency is determined from plots of $1/\xi_{\text{FMR}}$ versus $1/t$, as described in the main text. Here, there is only negligible Oersted field due to the in-plane current flow in the extremely resistive thin Ti layer [S7]. As shown in Fig. S13(b), $4\pi M_{\text{eff}}$ can be determined from the resonance frequency ($f$) dependence of $H_r$ following the Kittel's equation [S6]

$$f = \frac{\gamma}{2\pi} \sqrt{(H_r + H_{\text{ex}})(H_r + H_{\text{ex}} + 4\pi M_{\text{eff}})}, \quad (S13)$$

where $H_{\text{ex}}$ is the exchange field from the other layer.

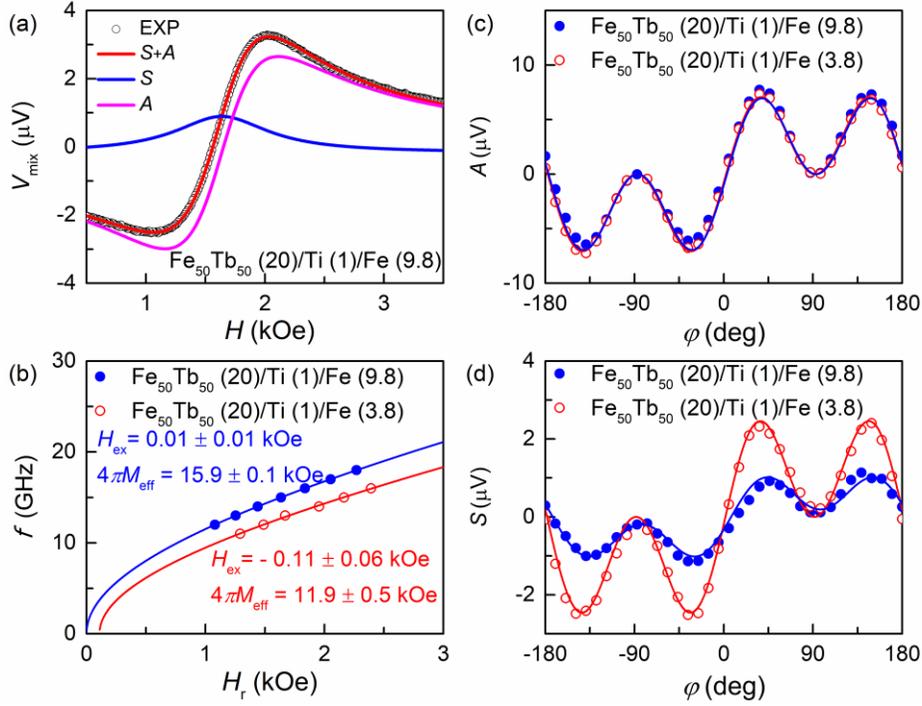

**Figure S13.** (a) ST-FMR spectrum at 15 GHz for $Fe_{50}Tb_{50}$ (20 nm)/Ti (1 nm)/Fe ($t_{Fe}$), in which a clear symmetric component (solid blue line) is from the damping-like torque, and the asymmetric component (solid pink line) is from the field-like torque and Oersted field torque. (b) Frequency dependence of ferromagnetic resonance field $H_r$ of the Fe layer in the representative samples $Fe_{50}Tb_{50}$ (20 nm)/Ti (1 nm)/Fe (9.8 nm) and $Fe_{50}Tb_{50}$ (20 nm)/Ti (1 nm)/Fe (3.8 nm). The solid red line represents fit of the data to Kittel's equation (Eq. (13)). (c) $A$ and (d) $S$ of the representative samples $Fe_{50}Tb_{50}$ (20 nm)/Ti (1 nm)/Fe (9.8 nm) (12 GHz) and $Fe_{50}Tb_{50}$ (20 nm)/Ti (1 nm)/Fe (3.8 nm) (14 GHz) plotted as a function of in-plane angle ($\varphi$) of the magnetic field relative to the rf current. The blue and red solid curves represent the best fits of the data to a $\sin2\varphi \cos\varphi$ dependence.

**Section 14. Interfacial perpendicular magnetic anisotropy energy density**

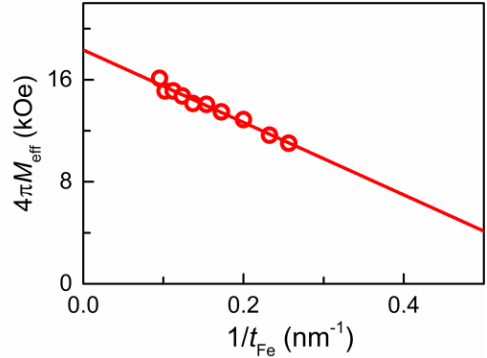

**Figure S14.** Dependence of the effective demagnetization field ($4\pi M_{eff}$) on the inverse thickness of the Fe layer in the $Fe_{50}Tb_{50}$ (20 nm)/Ti(1 nm)/Fe($t_{Fe}$) sample. The linear fit of the data to the relation $4\pi M_{eff} \approx 4\pi M_s - 2K_s/M_s t_{FM}$ yields a total interfacial magnetic anisotropy energy density of $K_s = 2.07 \pm 0.05$ erg/cm$^2$ and a saturation magnetization $M_s = 1500$ emu/cm$^3$. After subtraction of the contribution from the top Fe/MgO interface ($\approx 0.64$ erg/cm$^2$) [S8], we obtain the $K_s \approx 1.43 \pm 0.05$ erg/cm$^2$ for the Ti/Fe interface.

**Section 15. HHVR measurement of a 20 nm thick composition-gradient Fe$_{100-x}$Tb$_x$ ($x$ = 27→41)**

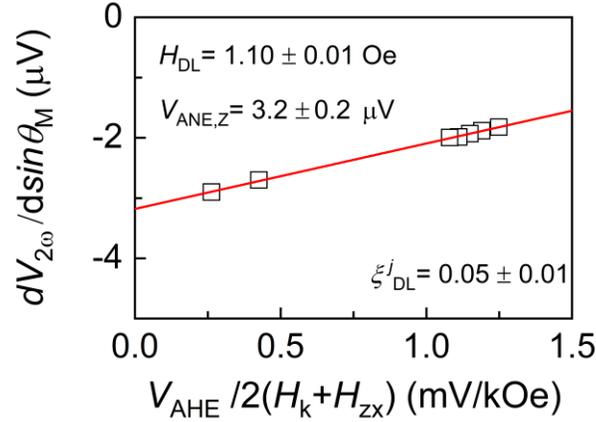

**Figure S15** $dV_{2\omega}/d\sin\theta_M$ versus $V_{AHE}/2(H_k+H_{zx})$ for the 20 nm thick composition-gradient Fe$_{100-x}$Tb$_x$ ($x$ = 27→41) sample, where $x$ is varied continuously from 27 at the bottom of the layer to 41 at the top. Using the extracted value of $H_{DL}$ obtained from a linear fit to Eq. (3) in the main text, we obtain $\xi_{DL}^{j}$ = 0.05 ±0.01 for this sample.

**Section 16. More examples of magnetization switching by the bulk spin-orbit torque**

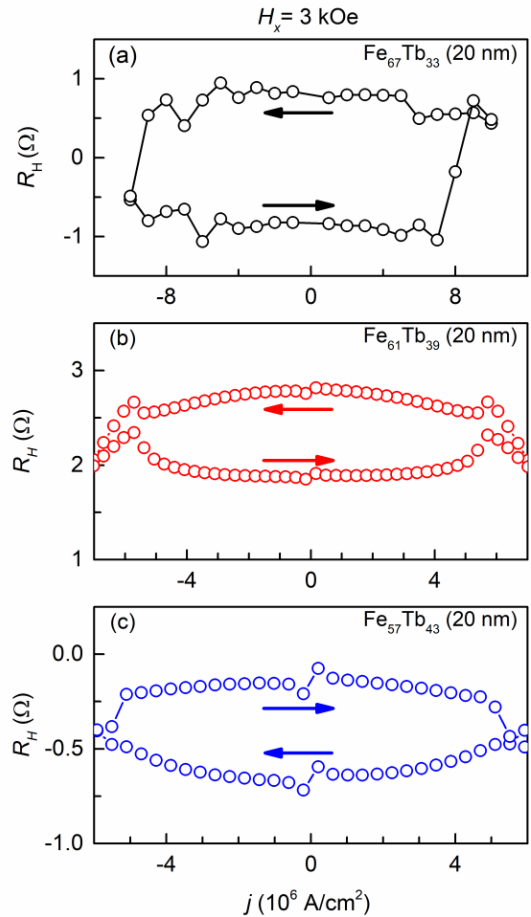

**Figure S16**. Current-induced switching of the 20 nm Fe$_{67}$Tb$_{33}$, Fe$_{61}$Tb$_{39}$, and Fe$_{57}$Tb$_{43}$ films under the same applied bias magnetic field of +3 kOe. The arrows indicate the switching polarity.

In Fig. S16(a)-(c), we show the current-induced switching of the 20 nm-thick $Fe_{67}Tb_{33}$, $Fe_{61}Tb_{39}$, and $Fe_{57}Tb_{43}$, under the same applied bias magnetic field of 3 kOe. Since all the three films are Fe-dominated ($R_{AHE} > 0$), their anomalous Hall voltages have the same sign. As indicated by the arrows, the switching polarity of the $Fe_{67}Tb_{33}$ is opposite to that of $Fe_{57}Tb_{43}$, which is consistent with the opposite sign of the bulk spin-orbit torques in the $Fe_{67}Tb_{33}$ ($\xi_{DL}^{j} > 0$) and the $Fe_{57}Tb_{43}$ ($\xi_{DL}^{j} < 0$). Interesting, the $Fe_{61}Tb_{39}$ shows the same switching polarity as the $Fe_{67}Tb_{33}$, which is because the current-induced heating during the switching has driven the $Fe_{61}Tb_{39}$ deep into the Fe-dominated regime where the $Fe_{67}Tb_{33}$ is also located and the bulk spin-orbit torque efficiency is positive ($\xi_{DL}^{j} > 0$). It has been reported that the ferrimagnetic CoTb near the compensation point can become Co-dominated at higher temperatures and Tb dominated at lower temperatures [S9].